\documentclass[a4paper,12pt]{article}
\pdfoutput=1
\usepackage{xcolor}
\usepackage{enumerate}
\usepackage{framed}
\usepackage{mdframed}
\usepackage{paralist}
 \mdfsetup{%
 linecolor=white,
 backgroundcolor=gray!20,
 }
\usepackage{amsmath}
\usepackage{amsfonts}
\usepackage{mathrsfs}
\usepackage{amssymb}
\usepackage[utf8x]{inputenc}
\usepackage{graphicx, rotating}
\usepackage{epsfig}
\usepackage{latexsym}
\usepackage{graphicx}
\usepackage{color}
\usepackage{amsmath,bm,amssymb}
\usepackage{cite}
\usepackage{slashed}
\usepackage{hyperref}
\usepackage{epstopdf}
\usepackage{orcidlink}
\usepackage[font=footnotesize,labelfont=]{caption}
\hypersetup{colorlinks=true, citecolor=bluscuro, linkcolor=black, urlcolor=bluscuro}
\definecolor{rossos}{cmyk}{0,1,1,0.55}
\definecolor{bluscuro}{rgb}{0.15, 0.2, .85}
\definecolor{bluchiaro}{cmyk}{1,.3,0.,0.1}
\definecolor{Gray}{gray}{0.95}

\def\to{\rightarrow}
\def\bea{\begin{eqnarray}}
\def\eea{\end{eqnarray}}

 \def\be   {\begin{equation}}   \def\ee   {\end{equation}}
 \def\ba   {\begin{array}}      \def\ea   {\end{array}}

\def\wh{\widehat{h}}

\newcommand{\Hub}{H}

\def\circa#1{\,\raise.3ex\hbox{$#1$\kern-.75em\lower1ex\hbox{$\sim$}}\,}

\newcommand{\beq}{\begin{equation}}
\newcommand{\eeq}{\end{equation}}

\font\tenrsfs=rsfs10 at 12pt
\font\sevenrsfs=rsfs7
\font\fiversfs=rsfs5
\newfam\rsfsfam
\textfont\rsfsfam=\tenrsfs
\scriptfont\rsfsfam=\sevenrsfs
\scriptscriptfont\rsfsfam=\fiversfs
\def\mathscr#1{{\fam\rsfsfam\relax#1}}

\renewcommand\d{{\rm d}}

\def \gta {\mathrel{\vcenter
     {\hbox{$>$}\nointerlineskip\hbox{$\sim$}}}}

\begin{document}

\thispagestyle{empty}
\begin{flushright}
{\small
CERN-TH-2026-002
}
\end{flushright}

\vspace{0.1cm}
\begin{center}
{\Large \bf 
Standard Model Higgs  Peaks:\\
\vspace{0.4cm}
a Note on the Vacuum Instability during Inflation}  \\[20mm]

{\bf\large   
 G. Franciolini\orcidlink{0000-0002-6892-9145}$^{1,2}$}, {\bf\large  A. Kehagias\orcidlink{0000-0001-6080-6215
}$^{3}$} {\bf\large  and}  {\bf\large  
  A. Riotto\orcidlink{0000-0001-6948-0856}$^{4}$}  \\[5mm]
\vspace{0.2cm}

$^{1}${\it Dipartimento di Fisica e Astronomia “G. Galilei”, \\Università degli Studi di Padova, 
via Marzolo 8, I-35131 Padova, Italy}
\vspace{0.2cm}

$^{2}${\it INFN, Sezione di Padova, via Marzolo 8, I-35131 Padova, Italy}

\vspace{0.2cm}
$^{3}${\it  Physics Division, National Technical University of Athens, \\Athens, 15780, Greece}
\vspace{0.2cm}

$^{4}${\it  D\'epartement de Physique Th\'eorique   and Gravitational Wave Science Center\\
Universit\'e de Gen\`eve, 24 quai E. Ansermet, CH-1211 Geneva, Switzerland}

\vspace{18mm}
{\large \bf Abstract}
\begin{quote}
In the Standard Model, the Higgs potential develops an instability at high field values when the quartic self-coupling runs negative. Large quantum fluctuations during cosmic inflation could drive the Higgs field beyond the potential barrier, creating regions that would be catastrophic for our observable universe. We point out that the extreme-value statistics describing the peaks (maxima)  of  the Higgs values is the correct statistics to infer the  condition to avoid vacuum instability. Even if this statistics delivers a bound on the Hubble rate during inflation which is only a factor $\sqrt{2}$ stronger than the one commonly adopted   in the literature, it is qualitatively distinct and  we believe worthwhile  communicating it. 

\end{quote}
\end{center}

\vfill
\noindent\line(1,0){188}
{\scriptsize{ \\ E-mail:\texttt{  
\href{mailto:gabriele.franciolini@unipd.it}{gabriele.franciolini@unipd.it}, \texttt{  
\href{mailto:kehagias@central.ntua.gr}{kehagias@central.ntua.gr},
\texttt{  
\href{mailto:antonio.riotto@unige.ch}{antonio.riotto@unige.ch}
}}}

\pagebreak\large
\normalsize

\section{Introduction}
It has been recognized for some time that the experimentally determined values of the Standard Model (SM) Higgs boson mass and the top-quark mass lead to an intriguing and nontrivial implication: if the SM is assumed to be valid up to arbitrarily high energies with no additional degrees of freedom, the universe appears to reside extremely close to the critical boundary between absolute stability and metastability of the electroweak (EW) vacuum (see, e.g., Refs.~\cite{Ellis:2009tp, Elias-Miro:2011sqh,
Degrassi:2012ry,
Buttazzo:2013uya} for recent studies).
More specifically, the Higgs effective potential within the SM develops an instability well below the Planck scale, at an energy scale of order $h_{\rm max} \sim 10^{12}$ GeV (for a gauge-invariant definition of this scale, see Ref.~\cite{Espinosa:2016nld}). Nevertheless, the decay time of the present electroweak vacuum is vastly longer than the age of the universe, making it effectively stable for all practical cosmological purposes.

This conclusion, however, does not necessarily extend to the early universe. At very early times, the presence of a field direction that is unbounded from below and corresponds to negative potential energy may severely threaten vacuum stability. A notable example arises during cosmic inflation \cite{Lyth:1998xn}. If the Higgs field is minimally coupled to gravity and does not interact directly with the inflaton, it behaves effectively as a massless scalar. In such a situation, quantum fluctuations  can become large enough to push the Higgs field away from the electroweak minimum and over the potential barrier,  as first pointed out in Ref.  \cite{Espinosa:2007qp} (see also Refs.~\cite{Herranen:2014cua,Hook:2014uia,Espinosa:2015qea,Franciolini:2018ebs} and the review~\cite{Markkanen:2018pdo}).
Should this occur, inflation would generate roughly spherical patches where the Higgs field reaches large values and explores the unstable region of the potential. These patches correspond to anti-de Sitter domains, which would be disastrous: after inflation ends, regions with negative vacuum energy expand at nearly the speed of light and eventually engulf the entire universe \cite{Espinosa:2015qea}. Therefore, consistency demands that the probability of producing an expanding anti-de Sitter bubble within our past light-cone be exceedingly small (see also Refs.~\cite{DeLuca:2022cus,Strumia:2022kez}).

A practical framework for tracking the evolution of Higgs fluctuations during inflation is the stochastic formalism based on a Fokker-Planck equation. This equation governs the probability distribution $P(h,N)$ for the Higgs field to take a value $h$ after $N$ e-folds of inflation (see, for instance, Ref. \cite{Linde:1990flp}),

\beq
\displaystyle \frac{\partial P}{\partial N} = \frac{\partial^2 }{\partial h^2} \bigg(  \frac{\Hub^2}{8\pi^2}  P\bigg)+
\frac{\partial }{\partial h}  \bigg(\frac{V'}{3\Hub^2} P\bigg),
\label{Fokker}
\eeq
where $H$ denotes the Hubble rate during inflation. Near the top of the potential barrier, the SM Higgs potential can be well approximated analytically by
\be
V(h) \approx - b \ln \left(\frac{ h^2}{h_{\rm max}^2 \sqrt{e}}\right) \frac{ h^4}{4}.
\label{StrV}
\ee
Here  $b \approx 0.16/(4\pi)^2\simeq 10^{-3}$ corresponds to the value of the $\beta$ function of the quartic coupling $\lambda$ evaluated around $h_{\rm max}$.

As shown  numerically in Ref.~\cite{Espinosa:2015qea}, because the Higgs quartic coupling vanishes close to the instability scale, the classical drift driven by the potential gradient is negligible over a wide range of field values compared to quantum fluctuations. Consequently, and  even if the Higgs field starts at $h=0$, its distribution rapidly approaches a Gaussian with zero mean and a variance that increases with the (square root of the) number of e-folds,
\beq\label{eq:hGauss}
P(h,N) = \frac{1}{\sqrt{2\pi \sigma_h^2}}\exp \left( -\frac{h^2}{2\sigma_h^2}\right) ,  ~~~~~
\sigma_h = \frac{\Hub}{2\pi} \sqrt{N}.\eeq
The  probability that, after $N$ e-folds, the Higgs field exceeds the top of the barrier is therefore
\beq \label{phmax}
p(h> h_{\rm max}) =\int_{h_{\rm max}}^\infty{\rm d} h\, P(h,N)\approx\frac{1}{2} \hbox{Erfc}\bigg(\frac{\sqrt{2}\pi  h_{\rm max}}{\sqrt{N} \Hub}\bigg).\eeq
Such  probability must be smaller than $e^{-3 N}$ in order to avoid finding such a region within our observable universe. This requirement follows from the fact that inflation produces
\be
{\cal N}= e^{3N}
\ee
causally disconnected Hubble-sized patches, each of which contributes to the present observable universe. Using the large-$x$ approximation $\hbox{Erfc}(x) \simeq e^{-x^2}/\sqrt{\pi}x$ and taking $N\simeq 60$ (to reproduce our observable universe), one finds (neglecting the prefactor does not change the numerics) \cite{Espinosa:2015qea}
\beq 
{\rm exp}\left(-2\pi^2h_{\rm max}^2/H^2N\right)< {\rm exp}\left(-3N\right)
\qquad{\rm or}\qquad
\frac{\Hub}{h_{\rm max}} < \sqrt{\frac23} \frac{\pi}{N}  \approx 0.04.
\label{box1}
\eeq
The purpose of this short note is to emphasize a  distinct point. Our observable universe is composed of roughly ${\cal N}$ independent regions generated during inflation. The presence of even a single region in which the Higgs field crosses the barrier and evolves toward the unstable side of the potential would be fatal. The upper bound (\ref{box1}) relies on the fact that each of these $\sim e^{180}$ regions corresponds to a realization of a nearly normally distributed random variable. However, away from the bulk of the distribution and on the tail of it, deviations from perfect Gaussian behavior can and do occur.  Although such events -- outliers --  are rare, they are precisely the ones of interest here, since their occurrence could jeopardize our very existence. In particular, what about   rare Higgs field maxima, or peaks?

The central question we address is the likelihood that the Higgs field accesses the unstable region due to the presence of such peaks. Answering this requires going beyond the statistics of averages and instead invoking 
extreme-value statistics, also known as supreme statistics \cite{bhavsar1985firstranked,coles2001introduction}, which describe the distribution of the largest value in a sample. The maximum Higgs excursion among all regions is the relevant quantity that determines whether the SM vacuum remains cosmologically viable during inflation. We discuss it in the next section.

\section{Higgs peaks and conclusions}

Consider a total of ${\cal N}$ elements drawn from the same underlying distribution, partitioned into ${\cal N}/m$ independent samples, each sample consisting of $m$ elements. The distribution of the maximum within each sample, and hence the most probable value of the overall maximum, can be derived using supreme statistics. In the limit ${\cal N}\to\infty$, one trivially recovers the fact that the maximum diverges.

In the present context, we have $m=1$ and ${\cal N}=e^{3N}\sim e^{180}$. For a Gaussian parent distribution, the probability that the Higgs field in a single comoving Hubble patch is smaller than a given value $\wh$ is described by the cumulative distribution function
\begin{eqnarray}
P_1(\wh) = \int_{-\infty}^{\wh} \d h P(h) 
=1-\frac{1}{2}{\rm Erfc}\left(\frac{\wh}{\sqrt{2}\sigma_h}\right).
\label{p1}
\end{eqnarray}
The probability that all ${\cal N}$ uncorrelated regions have Higgs field values below $\wh$ is then
\begin{equation}
\label{eq:peak_prof}
P_{\cal N}(\wh) = \left[P_1(\wh)\right]^{\cal N},
\end{equation}
and the probability density that a region with value $\wh$ represents the largest peak is given by $\d P_{\cal N}/\d \wh$. From this distribution one can compute both the mean and variance of the highest Higgs peaks.

In the regime of large Higgs field values, a simple analytic estimate for the most likely maximum can be easily obtained fluctuation \cite{MoradinezhadDizgah:2013rkr,MoradinezhadDizgah:2019wjf} (see also Ref. \cite{Dalal:2010hy}). The maximum of $\d P_{\cal N}/\d \wh$ identifies the most probable value of the largest Higgs, 
\be
\frac{\d^2P_{\cal N}}{\d\wh^2}  =0 \qquad {\rm or} \qquad ({\cal N}-1) \left(\frac{\d P_1}{\d\wh} \right)^2 + P_1 \frac{\d^2 P_1}{\d \wh^2}=0.
\label{detbar}
\ee
Using $\d P_1/\d \wh=P(\wh)$ and noting that $P_1\simeq 1$ in the large-field limit, Eq.~\eqref{detbar} simplifies to
\be
({\cal N}-1) P^2(\wh) = -\frac{\d P(\wh)}{\d h},
\label{start}
\ee
which leads to
\be
\frac{{\cal N}-1}{\sqrt{2\pi}} e^{-\wh^2/2\sigma_h^2} = \frac{\wh}{\sigma_h}.
\ee
Since ${\cal N}\ggg 1$, one finds that typical Higgs maximum is at 
\be\label{avde}
\wh\simeq \sqrt{2} \sigma_h\ln^{1/2} {\cal N}\simeq \sqrt{\frac{3}{2}}\frac{H}{\pi}N,
\ee
which diverges as $N\to\infty$, as expected and is larger by a factor $\sqrt{6\cdot 60}\approx 19$ than the square root of the variance $\sigma_h$. A more precise treatment in the large-${\cal N}$ limit shows that the probability distribution for the largest Higgs value is given by \cite{bhavsar1985firstranked,coles2001introduction}
\be\label{dalalpdf}
P_{\cal N}(\wh)=\frac{\d}{\d\wh}
\left\{\exp\left[-\frac{{\cal N}}{\sqrt{2\pi}}\frac{\sigma_h}{\wh}
e^{-\wh^2/2\sigma_h^2}
\right]\right\},
\ee
which is properly normalized to unity for $0<\wh<\infty$, see Fig.~\ref{fig:1}.  The double exponential is not specific of the Gaussian statistics, but it arises in the extreme-value statistics for any parent distribution which is sufficiently steep \cite{bhavsar1985firstranked}. The
probability that the Higgs peak exceeds the maximum of the SM potential is therefore
\beq \label{a}
p(\wh> h_{\rm max}) =\int_{h_{\rm max}}^\infty{\rm d} \wh P_{\cal N}(\wh)\approx\frac{{\cal N}\sigma_h}{\sqrt{2\pi}h_{\rm max}} e^{-h_{\rm max}^2/2\sigma_h^2}=
\frac{e^{3N}H\sqrt{N}}{(2\pi)^{3/2}h_{\rm max}} e^{-2\pi^2 h_{\rm max}^2/H^2 N}.
\eeq
Requiring this probability to be smaller than $e^{-3N}$ yields
\beq 
{\rm exp}\left(3N-2\pi^2h_{\rm max}^2/H^2N\right)< {\rm exp}\left(-3N\right)\qquad{\rm or}\qquad
\frac{\Hub}{h_{\rm max}} < \frac{1}{\sqrt{3}} \frac{\pi}{N}  \approx 0.03.
\label{boxmax}
\eeq
Parametrically, this constraint has the same $N$-dependence as Eq.~(\ref{box1}), but numerically is stronger by a factor of $\sqrt{2}$. More importantly, the bound in Eq.~(\ref{boxmax}) is conceptually distinct, as it is derived from the statistics of extreme values rather than from average behavior. 

\begin{figure}[t]\centering
\includegraphics[width=0.6 \textwidth]{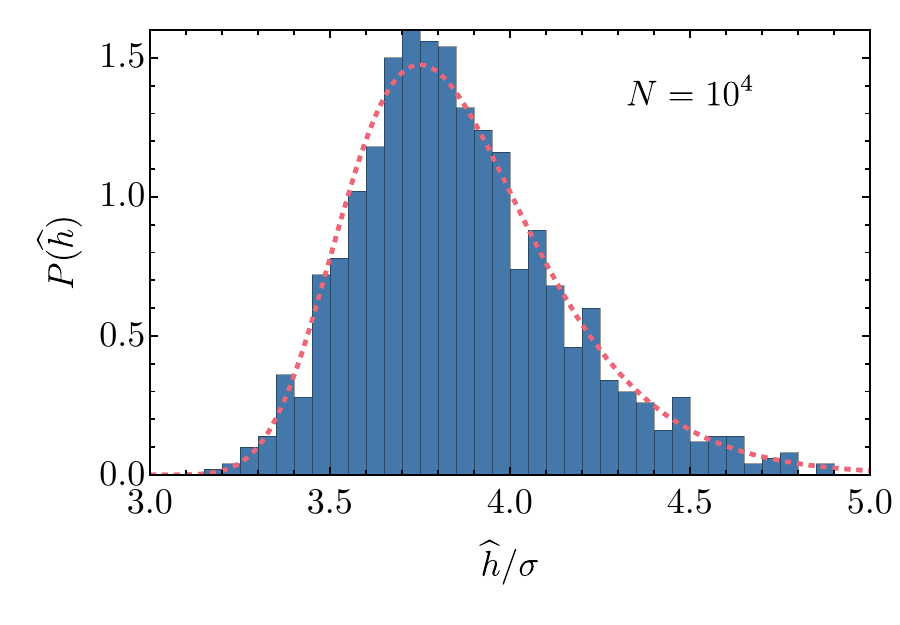}
\caption{The probability distribution for the extreme value of $h$ assuming $N=10^4$ realizations. We compare $10^3$ Monte Carlo realizations and the analytical distribution.}
\label{fig:1}
\end{figure}

Throughout this analysis we have assumed Gaussian statistics for the Higgs fluctuations, following Ref.~\cite{Espinosa:2015qea}, which is justified by the dominance of quantum effects near the instability scale. If instead one considers the opposite regime in which classical evolution dominates, the probability distribution becomes independent of $N$, $P(h)\sim {\rm exp}(-8\pi^2 V(h)/3H^4)$ \cite{Espinosa:2007qp} (the same result is obtained considering the Hawking-Moss bounce \cite{Joti:2017fwe}). Applying extreme-value statistics in that case leads to the condition
\be
{\rm exp}\left(3N-8\pi^2V(h_{\rm max})/3 H^4\right)< {\rm exp}\left(-3N\right),
\ee
which differs from the corresponding bound obtained from ensemble averages by a factor  $2^{1/4}$. 

Similarly, and maybe more interestingly,  one may consider the coupling with the curvature scalar $\sim \xi R h^2$ which induces a squared mass to the Higgs $\sim -12\xi H^2$. For negative and sufficiently small $\xi$ so that the Higgs is light enough to be quantum-mechanically excited during inflation ($\xi>-3/16$), neglecting the small Higgs quartic coupling, the Higgs probability is again a Gaussian of the form (\ref{eq:hGauss}), but with square root of the variance which quickly reaches the asymptotic value \cite{Espinosa:2015qea}
\be
\sigma_h=\frac{H}{4\pi\sqrt{-2\xi}}.
\ee
The extreme-value statistics will therefore lead to a bound again a factor of $\sqrt{2}$ stronger than what found in Ref. \cite{Espinosa:2015qea}
\be
\frac{\Hub}{h_{\rm max}} < 4\pi\sqrt{\frac{-2\xi}{3N}}.
\ee
For positive  and sufficiently large couplings, $\xi\gta 10^{-1}$,  the Higgs dynamics is dominated by classical evolution. Either the Higgs
is frozen during inflation at a new minimum and protected from quantum fluctuations, or is
driven classically to the dangerous AdS region \cite{Herranen:2014cua,Espinosa:2015qea,Franciolini:2018ebs}.

As a final point, one might wonder what is the impact of the second, third, $\cdots$, $q-$th maxima, 
 which are separated on average by a tiny amount \cite{MoradinezhadDizgah:2019wjf}

\be
\wh_{q+1}-\wh_q\simeq \frac{\sigma_h}{\log_{10} {\cal N}}\simeq \frac{\sigma_h}{\log_{10} e^{180}}\simeq \frac{\sigma_h}{78}
\ee
and 
whose probabilities might be in principle  larger than the one of the first maximum.  However, and again for a Gaussian parent distribution,  such probabilities satisfy the hierarchy \cite{MoradinezhadDizgah:2019wjf}

\be
\frac{P_{\cal N}(q+1,\wh) }{P_{\cal N}(q,\wh)}\simeq \frac{({\cal N}-q)}{2q}\,{\rm Erfc}\left(\frac{\wh}{\sqrt{2}\sigma_h}\right),
\ee
which implies

\be
p(q,\wh> h_{\rm max})\simeq \left[p(1,\wh> h_{\rm max})\right]^q < p(1,\wh> h_{\rm max}). 
\ee
We may safely  conclude that the role of the other  maxima is  negligible.  

\subsubsection*{Acknowledgments}
A.R.  acknowledges support from the  Swiss National Science Foundation (project number CRSII5\_213497).

\vskip 1cm

\bibliographystyle{JHEP}
\bibliography{draft}

@book{Linde:1990flp,
    author = "Linde, Andrei D.",
    title = "{Particle physics and inflationary cosmology}",
    eprint = "hep-th/0503203",
    archivePrefix = "arXiv",
    volume = "5",
    year = "1990"
}

@article{Dalal:2010hy,
    author = "Dalal, Neal and Lithwick, Yoram and Kuhlen, Michael",
    title = "{The Origin of Dark Matter Halo Profiles}",
    eprint = "1010.2539",
    archivePrefix = "arXiv",
    primaryClass = "astro-ph.CO",
    month = "10",
    year = "2010"
}

@article{DeLuca:2022cus,
    author = "De Luca, Valerio and Kehagias, Alex and Riotto, Antonio",
    title = "{On the cosmological stability of the Higgs instability}",
    eprint = "2205.10240",
    archivePrefix = "arXiv",
    primaryClass = "hep-ph",
    doi = "10.1088/1475-7516/2022/09/055",
    journal = "JCAP",
    volume = "09",
    pages = "055",
    year = "2022"
}

@article{MoradinezhadDizgah:2019wjf,
    author = "Moradinezhad Dizgah, Azadeh and Franciolini, Gabriele and Riotto, Antonio",
    title = "{Primordial Black Holes from Broad Spectra: Abundance and Clustering}",
    eprint = "1906.08978",
    archivePrefix = "arXiv",
    primaryClass = "astro-ph.CO",
    doi = "10.1088/1475-7516/2019/11/001",
    journal = "JCAP",
    volume = "11",
    pages = "001",
    year = "2019"
}

@article{MoradinezhadDizgah:2013rkr,
    author = "Moradinezhad Dizgah, Azadeh and Dodelson, Scott and Riotto, Antonio",
    title = "{Imprint of Primordial Non-Gaussianity on Dark Matter Halo Profiles}",
    eprint = "1307.2632",
    archivePrefix = "arXiv",
    primaryClass = "astro-ph.CO",
    reportNumber = "FERMILAB-PUB-13-275-A",
    doi = "10.1103/PhysRevD.88.063513",
    journal = "Phys. Rev. D",
    volume = "88",
    pages = "063513",
    year = "2013"
}

@article{Joti:2017fwe,
    author = "Joti, Aris and Katsis, Aris and Loupas, Dimitris and Salvio, Alberto and Strumia, Alessandro and Tetradis, Nikolaos and Urbano, Alfredo",
    title = "{(Higgs) vacuum decay during inflation}",
    eprint = "1706.00792",
    archivePrefix = "arXiv",
    primaryClass = "hep-ph",
    reportNumber = "CERN-TH-2017-121",
    doi = "10.1007/JHEP07(2017)058",
    journal = "JHEP",
    volume = "07",
    pages = "058",
    year = "2017"
}

@article{bhavsar1985firstranked,
  author  = {Bhavsar, S.\,P. and Barrow, J.\,D.},
  title   = {First ranked galaxies in groups and clusters},
  journal = {Monthly Notices of the Royal Astronomical Society},
  volume  = {213},
  number  = {4},
  pages   = {857--869},
  year    = {1985},
  doi     = {10.1093/mnras/213.4.857}
}

@book{coles2001introduction,
  title     = {An Introduction to Statistical Modeling of Extreme Values},
  author    = {Coles, Stuart},
  series    = {Springer Series in Statistics},
  publisher = {Springer London},
  year      = {2001},
  isbn      = {978-1-85233-459-8, 978-1-84996-874-4},
  doi       = {10.1007/978-1-4471-3675-0}
}

@article{Strumia:2022kez,
    author = "Strumia, Alessandro and Tetradis, Nikolaos",
    title = "{Higgstory repeats itself}",
    eprint = "2207.00299",
    archivePrefix = "arXiv",
    primaryClass = "hep-ph",
    doi = "10.1007/JHEP09(2022)203",
    journal = "JHEP",
    volume = "09",
    pages = "203",
    year = "2022"
}

@article{Franciolini:2018ebs,
    author = "Franciolini, G. and Giudice, G. F. and Racco, D. and Riotto, A.",
    title = "{Implications of the detection of primordial gravitational waves for the Standard Model}",
    eprint = "1811.08118",
    archivePrefix = "arXiv",
    primaryClass = "hep-ph",
    reportNumber = "CERN-TH-2018-233",
    doi = "10.1088/1475-7516/2019/05/022",
    journal = "JCAP",
    volume = "05",
    pages = "022",
    year = "2019"
}

@article{Ellis:2009tp,
    author = "Ellis, J. and Espinosa, J. R. and Giudice, G. F. and Hoecker, A. and Riotto, A.",
    title = "{The Probable Fate of the Standard Model}",
    eprint = "0906.0954",
    archivePrefix = "arXiv",
    primaryClass = "hep-ph",
    reportNumber = "CERN-PH-TH-2009-058",
    doi = "10.1016/j.physletb.2009.07.054",
    journal = "Phys. Lett. B",
    volume = "679",
    pages = "369--375",
    year = "2009"
}

@article{Buttazzo:2013uya,
    author = "Buttazzo, Dario and Degrassi, Giuseppe and Giardino, Pier Paolo and Giudice, Gian F. and Sala, Filippo and Salvio, Alberto and Strumia, Alessandro",
    title = "{Investigating the near-criticality of the Higgs boson}",
    eprint = "1307.3536",
    archivePrefix = "arXiv",
    primaryClass = "hep-ph",
    reportNumber = "CERN-PH-TH-2013-166, FTUAM-13-20, IFT-UAM-CSIC-13-081, IFUP-TH",
    doi = "10.1007/JHEP12(2013)089",
    journal = "JHEP",
    volume = "12",
    pages = "089",
    year = "2013"
}

@article{Espinosa:2016nld,
    author = "Espinosa, Jose R. and Garny, Mathias and Konstandin, Thomas and Riotto, Antonio",
    title = "{Gauge-Independent Scales Related to the Standard Model Vacuum Instability}",
    eprint = "1608.06765",
    archivePrefix = "arXiv",
    primaryClass = "hep-ph",
    reportNumber = "DESY-16-161, CERN-TH-2016-187",
    doi = "10.1103/PhysRevD.95.056004",
    journal = "Phys. Rev. D",
    volume = "95",
    number = "5",
    pages = "056004",
    year = "2017"
}

@article{Markkanen:2018pdo,
    author = "Markkanen, Tommi and Rajantie, Arttu and Stopyra, Stephen",
    title = "{Cosmological Aspects of Higgs Vacuum Metastability}",
    eprint = "1809.06923",
    archivePrefix = "arXiv",
    primaryClass = "astro-ph.CO",
    reportNumber = "IMPERIAL/TP/2018/TM/05",
    doi = "10.3389/fspas.2018.00040",
    journal = "Front. Astron. Space Sci.",
    volume = "5",
    pages = "40",
    year = "2018"
}

@article{Hook:2014uia,
    author = "Hook, Anson and Kearney, John and Shakya, Bibhushan and Zurek, Kathryn M.",
    title = "{Probable or Improbable Universe? Correlating Electroweak Vacuum Instability with the Scale of Inflation}",
    eprint = "1404.5953",
    archivePrefix = "arXiv",
    primaryClass = "hep-ph",
    reportNumber = "MCTP-14-10",
    doi = "10.1007/JHEP01(2015)061",
    journal = "JHEP",
    volume = "01",
    pages = "061",
    year = "2015"
}

@article{Lyth:1998xn,
    author = "Lyth, David H. and Riotto, Antonio",
    title = "{Particle physics models of inflation and the cosmological density perturbation}",
    eprint = "hep-ph/9807278",
    archivePrefix = "arXiv",
    reportNumber = "LANCS-TH-9720, FERMILAB-PUB-97-292-A, CERN-TH-97-383, OUTP-98-39-P",
    doi = "10.1016/S0370-1573(98)00128-8",
    journal = "Phys. Rept.",
    volume = "314",
    pages = "1--146",
    year = "1999"
}

@article{Espinosa:2015qea,
    author = "Espinosa, Jose R. and Giudice, Gian F. and Morgante, Enrico and Riotto, Antonio and Senatore, Leonardo and Strumia, Alessandro and Tetradis, Nikolaos",
    title = "{The cosmological Higgstory of the vacuum instability}",
    eprint = "1505.04825",
    archivePrefix = "arXiv",
    primaryClass = "hep-ph",
    reportNumber = "CERN-PH-TH-2015-119",
    doi = "10.1007/JHEP09(2015)174",
    journal = "JHEP",
    volume = "09",
    pages = "174",
    year = "2015"
}

@article{Degrassi:2012ry,
    author = "Degrassi, Giuseppe and Di Vita, Stefano and Elias-Miro, Joan and Espinosa, Jose R. and Giudice, Gian F. and Isidori, Gino and Strumia, Alessandro",
    title = "{Higgs mass and vacuum stability in the Standard Model at NNLO}",
    eprint = "1205.6497",
    archivePrefix = "arXiv",
    primaryClass = "hep-ph",
    reportNumber = "CERN-PH-TH-2012-134, RM3-TH-12-9",
    doi = "10.1007/JHEP08(2012)098",
    journal = "JHEP",
    volume = "08",
    pages = "098",
    year = "2012"
}

@article{Elias-Miro:2011sqh,
    author = "Elias-Miro, Joan and Espinosa, Jose R. and Giudice, Gian F. and Isidori, Gino and Riotto, Antonio and Strumia, Alessandro",
    title = "{Higgs mass implications on the stability of the electroweak vacuum}",
    eprint = "1112.3022",
    archivePrefix = "arXiv",
    primaryClass = "hep-ph",
    doi = "10.1016/j.physletb.2012.02.013",
    journal = "Phys. Lett. B",
    volume = "709",
    pages = "222--228",
    year = "2012"
}

@article{Espinosa:2007qp,
    author = "Espinosa, J. R. and Giudice, G. F. and Riotto, A.",
    title = "{Cosmological implications of the Higgs mass measurement}",
    eprint = "0710.2484",
    archivePrefix = "arXiv",
    primaryClass = "hep-ph",
    reportNumber = "CERN-PH-TH-2007-179, IFT-UAM-CSIC-07-50",
    doi = "10.1088/1475-7516/2008/05/002",
    journal = "JCAP",
    volume = "05",
    pages = "002",
    year = "2008"
}

@article{Herranen:2014cua,
    author = "Herranen, Matti and Markkanen, Tommi and Nurmi, Sami and Rajantie, Arttu",
    title = "{Spacetime curvature and the Higgs stability during inflation}",
    eprint = "1407.3141",
    archivePrefix = "arXiv",
    primaryClass = "hep-ph",
    reportNumber = "IMPERIAL-TP-2014-AR-2",
    doi = "10.1103/PhysRevLett.113.211102",
    journal = "Phys. Rev. Lett.",
    volume = "113",
    number = "21",
    pages = "211102",
    year = "2014"
}

\end{document}